\documentclass[10pt]{iopart}
\usepackage{graphicx}

\begin{document}

\title[(001) and (111) surfaces of NiMnSb]
       {Structural and magnetic properties of the (001)
       and (111) surfaces of the half-metal NiMnSb}

\author{M Le\v{z}ai\'{c}\dag,
I Galanakis\ddag, G Bihlmayer\dag\ and S Bl\"ugel\dag}

\address{\dag\ Institut f\"ur Festk\"orperforschung,
Forschungszentrum J\"ulich, D-52425 J\"ulich, Germany}
\address{\ddag\ Institut of Microelectronics, NCSR ``Demokritos'',
15310 Aghia Paraskevi, Athens, Greece}

\ead{I.Galanakis@fz-juelich.de}

\begin{abstract}
Using the full potential linearised augmented planewave method we
study the electronic and magnetic properties of the (001) and
(111) surfaces of the half-metallic Heusler alloy NiMnSb from
first-principles. We take into account all possible surface
terminations including relaxations of these surfaces. Special
attention is paid to the spin-polarization at the Fermi level
which governs the spin-injection from such a metal into a
semiconductor. In general, these surfaces lose the half-metallic
character of the bulk NiMnSb, but for the (111) surfaces this loss
is more pronounced. Although structural optimization does not
change these features qualitatively, specifically for the (111)
surfaces relaxations can compensate much of the spin-polarization
at the Fermi surface that has been lost upon formation of the
surface.
\end{abstract}


\submitto{Journal of Physics: Condensed Matter}

\maketitle

\section{Introduction} \label{sec1}

During the last decade the emergence of the fields of
magnetoelectronics and spintronics has given birth to a new series
of challenges in materials science \cite{Zutic,deBoeck}. A central
problem remains the spin-injection from a ferromagnet into a
semiconductor \cite{Wunnicke}. Its successful realization would
lead to the creation of a series of novel devices such as
spin-filters \cite{Kilian}, tunnel junctions \cite{Tanaka} or GMR
devices for spin injection \cite{Caballero98}. The use of
half-metallic ferromagnets as electrodes was proposed to maximize
the efficiency of such spintronic devices.  These compounds are
ferromagnetic metals with a band gap at the Fermi level
($E_\mathrm{F}$) in the minority spin channel leading to 100\%
spin-polarization at $E_\mathrm{F}$. Thus, in principle, during
the spin-injection process only majority-spin electrons would be
injected in the semiconductor.

The family of half-metallic systems which has attracted most of
the attention are the half-Heusler alloys and especially NiMnSb.
These compounds of the general formula XYZ  crystallize in the
$C1_b$ structure, which consists of 4 fcc sublattices occupied by
the three atoms X, Y and Z and a vacant site \cite{Landolt}. In
1983 de Groot and his collaborators were the first to predict the
half-metallic character of NiMnSb on the basis of first-principles
calculations \cite{deGroot}. Thereafter, several ab-initio
calculations on NiMnSb reproduced these  results \cite{calc},
and Galanakis \textit{et al.}
showed that the gap arises from the hybridization between the $d$
orbitals of the Ni and Mn atoms \cite{GalanakisBulk}. This
explanation was confirmed also by the work of  Nanda and Dasgupta
\cite{Nanda}. Several other studies were performed on the
stability of the minority-spin band-gap which was found to be
stable under hydrostatic pressure and tetragonalization
\cite{Block} or a small disorder \cite{Orgassa} but the exchange
of the atoms occupying the different sublattices completely
destroys the gap \cite{Larson}. Experiments seem to well establish
the half-metallicity in the case of NiMnSb single crystals
\cite{Kirillova,Hanssen} or away from the surface in the case of
thick films \cite{Clowes}.

Recently, high quality films of NiMnSb alloys have been also
grown \cite{Roy,Gardelis,Bach}, but they were found to be
not half-metallic \cite{Soulen,films}; a maximum value of 58\% for
the spin-polarisation of NiMnSb was obtained by Soulen \textit{et
al.} \cite{Soulen}. These polarisation values are consistent with
a small perpendicular magnetoresistance measured for NiMnSb in a
spin-valve structure \cite{Caballero99}, and a superconducting and
a magnetoresistive tunnel junction \cite{Tanaka}. Ristoiu
\textit{et al.} showed that during the growth of the NiMnSb thin
films, first Sb and then Mn atoms segregate to the surface, which
is far from being perfect, thus decreasing the obtained
spin-polarization \cite{Ristoiu}. But when they removed the excess
of Sb by a flash annealing, they managed to get a nearly
stoichiometric ordered alloy surface being terminated by a MnSb
layer, which presented a spin-polarization of about 67$\pm$9\% at
room temperature \cite{Ristoiu}. Even in the case of perfect
surfaces, these would not be half-metallic due to surface states
\cite{GalanakisSurf001}, non-quasiparticle states \cite{Chioncel}
and finite-temperature effects \cite{Dowben}. There is evidence
that around 80 K the films undergo a transition towards a normal
metal \cite{Borca}.

First-principles calculations have been  also employed to study
the surfaces of NiMnSb.  Jenkins and King were the first to study
by a pseudopotential technique the MnSb terminated (001)
surface of NiMnSb and showed that there are two surface states at
the Fermi level, which are well localised in the surface layer
\cite{Jenkins01}. They have also shown that there is a small
relaxation of the surface with the Mn atoms moving slightly
inwards and the Sb outwards and this small relaxation is
energetically more favourable than the creation of Mn or Sb
dimers forming in a $c(2\times 2)$ reconstruction. Galanakis
studied also the (001) surfaces using the full-potential
version of the Korringa-Kohn-Rostoker Green function
method (FSKKR) \cite{GalanakisSurf001}. He found that
the MnSb-terminated surface shows a quite large spin-polarization
in agreement with the experiments of Ristoiu \textit{et al.}
\cite{Ristoiu}.

In a study of (111) terminated NiMnSb surfaces, Jenkins
investigated the relative stability of stoichiometric surfaces as
well as their stability with respect to to other,
nonstoichiometric structures \cite{Jenkins04}. The spin moments
and surface states of the (111) surfaces of NiMnSb have been
studied by Galanakis, who found in all cases very pronounced
surface states on the unrelaxed (111) surfaces
\cite{GalanakisSurf111}.  These calculations have been obtained
within the atomic sphere approximation which can lead to errors
for surfaces with respect to the full-potential ones.

In this communication, we present ab-initio calculations of the
(001) and (111) surfaces of the half-metallic NiMnSb Heusler
alloy. We take into account all possible terminations and study
the electronic and magnetic properties of the surfaces and
calculate the spin-polarization at the Fermi level. We also
investigate the effect of the relaxation of  the atomic positions
of the atoms near the surface on the electronic and magnetic
properties of the surfaces. Section \ref{sec2} presents the
method of the calculation. In section \ref{sec3} we briefly review
the bulk properties of NiMnSb and sections \ref{sec4} and
\ref{sec5} are devoted to the (001) and (111) surfaces,
respectively. Finally in section \ref{sec6} we resume and
conclude.

Prior to presenting our results we should make two important
notices. First, based on the experiences from ferromagnets and
semiconductors, two effects should be particularly relevant for
the surfaces of half-metals: (i) in magnetic systems, the moments
of the surface atoms are strongly enhanced due to the missing
hybridization with the cut-off neighbours, and (ii) in
semiconductors, surface states can appear in the gap such that the
surface often becomes metallic. Also this is a consequence of the
reduced hybridization, leading to dangling bond states in the gap.
Secondly, it should be mentioned that at an interface, the interface
states will certainly differ from the surface states studied here.
But to a certain extent, the surface states for a given surface
orientation will have characteristics also typical for interface
states.  In principle these states should not affect
the magnetoconductance since the wavefunction is orthogonal to all
bulk states incident to the surface. But emission or absorption of
magnons couples weakly the bulk and surface states and affects the
magnetotransport. In real systems the interaction of the surface
states with other defect  states in the bulk and/or with surface
defects makes the surface states conducting and leads to the low
spin-polarization values for films derived by Andreev reflection
measurements.

\section{Method and structure} \label{sec2}

The calculations were performed using density functional theory
and the generalized gradient approximation (GGA) as given by
Perdew et al. \cite{PBE}. We use the full-potential linearised
augmented planewave (FLAPW) method in film geometry \cite{Wimmer},
as implemented in the {\tt FLEUR} program \cite{FLEUR}. For the
calculations, a planewave cutoff $K_{\rm max}$ of 3.6~a.u.$^{-1}$
was used. Lattice harmonics with angular momentum $l\leq 8$ were
used to expand the charge density and the wavefunctions within the
muffin-tin spheres. In the case of the (001) surfaces we used a
film consisting of 9 atomic layers of NiMnSb, while for the (111)
interfaces 13 layers were used. The two-dimensional
Brillouin-zone was sampled with 64 special {\bf k}-points in the
irreducible wedge for the (001) surfaces, and 90 {\bf k}-points in
the irreducible wedge for the (111) surfaces. All the calculations
were performed at the optimized lattice constant of NiMnSb
(5.915~\AA) which is within 0.2\% agreement with the experimental
value \cite{Otto}. Structural optimization was done by minimizing
the forces on the three topmost layers of the film.

There are two different possible terminations in the case of the
(001) surfaces, one containing the Mn and Sb atoms while the other
contains only a Ni atom \cite{GalanakisSurf001,Jenkins01}. The
interlayer distance is one fourth of the bulk lattice constant. In the
perpendicular direction the layer occupancy is repeated every
fourth layer, since in the $i\pm 2$ layer the atoms have exchanged
positions compared to the $i$ layer.

\begin{figure}
\begin{center}
\includegraphics[angle=270,width=0.8\textwidth]{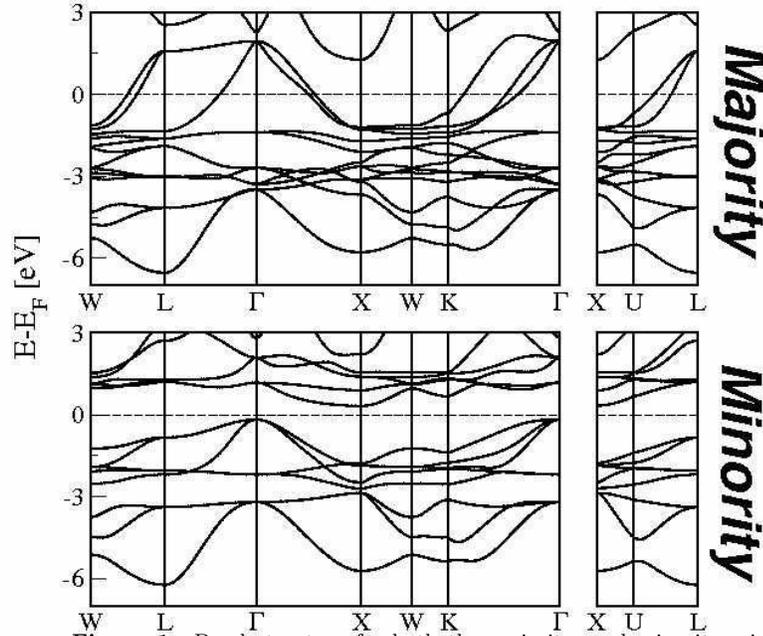}
\caption{Band structure for both the majority- and minority-spin
electrons along the high symmetry axis. The minority band is
semiconducting while the majority is metallic.} \label{fig1}
\end{center}
\end{figure}

In the case of the (111) surfaces we have more different surface
terminations as compared to the (001) ones. Along the [111]
direction the alloy consists of alternating closed packed layers
containing only one chemical element per layer. For example in the
case of a Ni-terminated surface, there are two different
possibilities: either to have a Mn subsurface layer or an Sb one
(see \cite{Jenkins04} or \cite{GalanakisSurf111} for the different
terminations). In total we can identify six different surface
terminations.

\section{Bulk properties} \label{sec3}

The bulk properties of NiMnSb have been extensively studied during
the last years and it is still considered to be a key component
in the search for spintronic devices. Thus, in this section we
will only briefly discuss its bulk properties. NiMnSb, as predicted
by de Groot and collaborators, is a half-metal \cite{deGroot}. To
explain this we present in figure~\ref{fig1} the band structure of
this compound for both majority- (upper panel) and minority-spin
(bottom panel) electrons. The majority-spin bands are
characteristic of a normal metal with the Fermi level crossing the
$d$-like bands. On the contrary the minority-spin bands are like in
a semiconductor and the Fermi level lies in a gap. Notice
that the gap is an indirect one between the $\Gamma$ and $X$
points in the  reciprocal b.c.c. unit cell.

\begin{figure}
\begin{center}
\includegraphics[angle=270,width=0.8\textwidth]{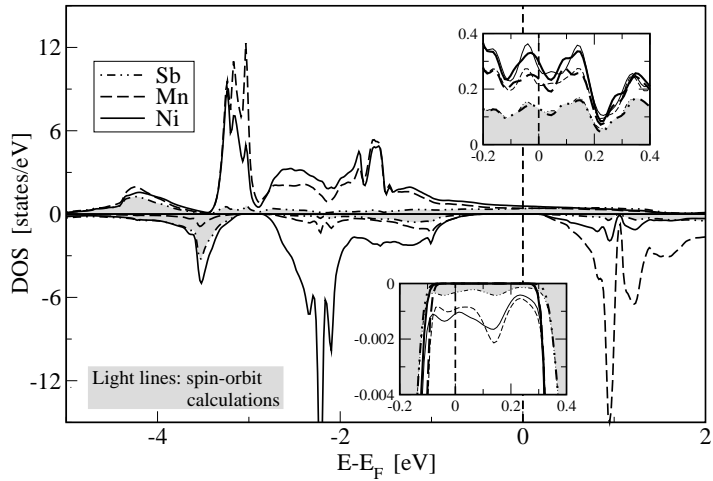}
\caption{Atom-projected density of states (DOS) of bulk NiMnSb.
The DOS in the interstitial region is not shown. The insets show
magnifications of the area around the gap; with the light lines we
indicate also the calculations where spin-orbit coupling has been
included. \label{fig2}}
\end{center}
\end{figure}

The character of each band has been extensively discussed in
reference \cite{GalanakisBulk}. The lower bands in the figure
\ref{fig1} arise from the $p$-states of Sb and the minority-spin
gap is created between the bonding and antibonding $d$-hybrids
created by the Mn and Ni atoms. The minority-spin bonding states
have most of their weight at the Ni atom and the antibonding at
the Mn atom leading to very large localised spin moments at the Mn
atoms \cite{Plogman}. There are exactly 9 occupied minority states
and the total spin moment follows the Slater-Pauling behaviour
shown in reference \cite{GalanakisBulk}, being exactly $4 \mu_B$.

In figure~\ref{fig2} we present  the atom-projected density of
states (DOS) which is defined by the muffin-tin spheres surrounding
each atom. We do not show the DOS for the interstitial region
(i.e.\ between the muffin-tin spheres) and the Sb
$s$-states low in energy. As we just discussed the lowest part of
the DOS are occupied by the $p$-states of Sb which couple also
to $p$-states at the other sites. Ni and Mn atoms create a common
majority $d$ band while, as we mentioned, the minority occupied
$d$-like hybrids are mainly located at the Ni atom leading to a
large Mn spin moment ($3.729 \mu_B$) and a small Ni one ($0.246
\mu_B$). The Sb moment is very small ($-0.066 \mu_B$) and
antiparallel to the spin moments of both Ni and Mn. Atomic spin
moments are obtained by integrating over the non-overlapping
muffin-tin spheres, so that the sum of the atomic moments is not
exactly $4 \mu_B$.

All calculations are done within the scalar-relativistic
approximation. Inclusion of spin-orbit coupling will couple the
two spin-directions and partially fill the minority gap.
Performing calculations that include  spin-orbit coupling
self-consistently, we find that the overall DOS and spin-moments
scarcely changed and thus we present the results only around the
gap in the insets of figure~\ref{fig2}. The majority-spin DOS
around the Fermi level changes only marginally while states are
now present within the gap. But the intensity of the minority-spin
DOS is two orders of magnitude smaller than the majority-spin DOS
at the Fermi level and instead of a gap there is now a region of
almost 100\% spin-polarization. These results agree with the work
presented in reference \cite{Mavropoulos}. The DOS induced by
minority-spin surface states will be much more important than the
contribution of  spin-orbit coupling, thus the latter quantity can
be neglected when studying the surface properties of NiMnSb. The
orbital magnetism in NiMnSb is discussed in reference
\cite{GalanakisOrbit}.

\begin{figure}
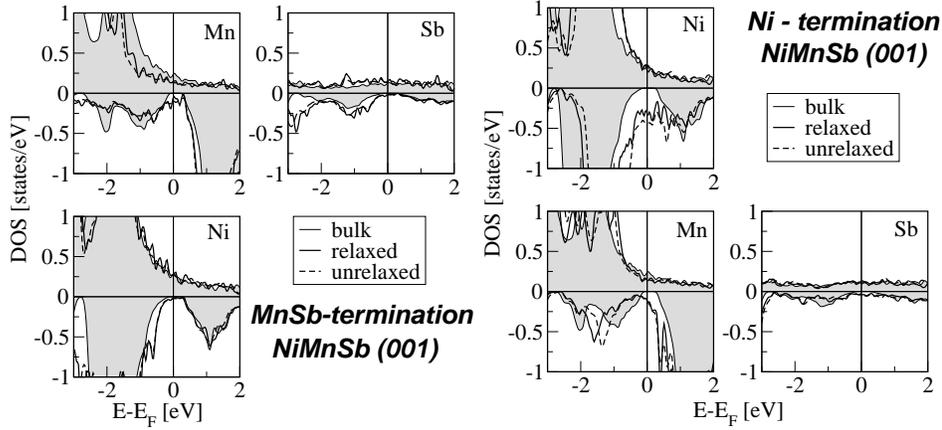

\begin{center}
\includegraphics[scale=0.33]{fig3a.eps}
\includegraphics[scale=0.33]{fig3b.eps}
\caption{Local DOS for the atoms at the surface and subsurface
layers for both Ni- and MnSb- terminated NiMnSb(001) surfaces.
The results of the relaxed and unrelaxed surfaces are indicated
by thick solid lines and  dashed lines, respectively. Grey shaded
regions  represent the bulk results.} \label{fig3}
\end{center}
\end{figure}

\section{(001) surfaces} \label{sec4}

\subsection{Structure and relaxation}

As described in section \ref{sec2}, there are two different terminations
for the (001) surfaces: a Ni or a MnSb layer. In both cases  the three top
layers were relaxed. In the case of the Ni termination, almost no buckling
or relaxation of the MnSb subsurface layer was observed, while the distance
between the top Ni layer and the  subsurface layer was reduced by
around 10\%. In the case of the MnSb termination,  the Mn atom at
the surface layer moves inwards and the Sb atom outwards. The distance
between the Mn surface atom and the Ni subsurface layer is contracted
by 3.5 \% and the distance between the Sb surface atom and the Ni
subsurface layer is expanded by 7.3 \%. Qualitatively, these results agree
with the results obtained for the MnSb termination of the (001)
surface by Jenkins and King \cite{Jenkins01}.

\subsection{Density of states and bandstructures}

In the right panel of figure~\ref{fig3} we present the atom- and
spin-projected densities of states for the Ni atom at the
surface and the Mn and Sb atoms in the subsurface  layer for the
case of the Ni-terminated surface. The left panel contains the
results for the MnSb terminated NiMnSb(001) surface. For both
possible terminations we include the surface DOS of both the
relaxed and unrelaxed calculations together with the bulk
results (grey region). We see that relaxation has a very small
effect on the DOS even around the Fermi level. In the following, only
the results including relaxation will be discussed.

\begin{figure}
\begin{center}
\includegraphics[scale=0.6]{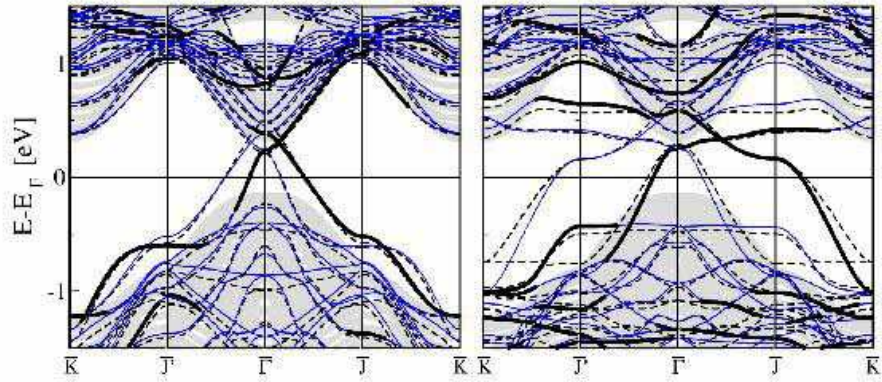}
\caption{Minority-spin surface bandstructure for the MnSb (left)
and Ni (right) terminated  (001) surfaces. Grey regions indicate
the projected  bulk bandstructure. Dashed lines show the
bandstructure for the unrelaxed surface, while thin full lines
indicate the result for the relaxed surface. Thick lines indicate
surface states on one of the of the two equivalent surfaces of the
film (we considered that a  surface state should have more than
50\% of its weight located at the first two surface layers).}
\label{fig4}
\end{center}
\end{figure}

In the case of the MnSb terminated surface, the DOS with the
exception of the gap area is very similar to the bulk
calculations. The Ni atom in the subsurface layer presents
practically half-metallic character with an almost zero
minority-spin DOS, while for the bulk there is an absolute gap.
The Mn and Sb atoms in the surface layer show more pronounced
differences with respect to the bulk, and within the gap  there is
a very small Mn-$d$ and Sb-$p$ DOS. These intensities are due to the
two surface states discussed already by Jenkins and King \cite{Jenkins01}.
These states are strongly localised  at the surface layer as in
the subsurface layer there are practically no states inside the
gap. This is in agreement with previous first principles
calculations by Galanakis \cite{GalanakisSurf001}. Our theoretical
results agree with the experiments of Ristoiu \textit{et al.}
\cite{Ristoiu} who in the case of a MnSb well ordered (001)
surface measured a high spin-polarization.

To examine the origin of these surface states we also calculated
the surface bandstructures shown in figure~\ref{fig4}. Here, the
two-dimensional Brillouin zone is a square. Relaxations (thin
solid lines) give rise to only small changes with respect to the
unrelaxed results (dashed lines). Since the films in our calculations
have two surfaces that are rotated by 90$^{\circ}$ with respect to
each other, the surface bandstructure seems to have a four-fold
symmetry axis through the $\overline{\Gamma}$ point. Thick
solid lines denote  surface states arising from only one
surface. Our results agree with the ones of Jenkins and King who
-- for the same surface -- have shown that there are two surface
states \cite{Jenkins01}. They noted that the lower lying state
(0.20 eV above the Fermi level, $E_\mathrm{F}$) is due to the
interaction between $e_g$-like dangling bond states located at the
Mn atoms, while the second surface state, ($\sim$0.3 eV above
$E_\mathrm{F}$) arises from the hybridization between
$t_{2g}$-like orbitals of Mn with $p$-type orbitals of Sb. The
first surface state disperses downwards along the
$\overline{\Gamma J}$ direction while the second surface state
disperses upwards along the same direction. Since their dispersion
reverses along the $\overline{\Gamma J'}$ direction, we assume
that there is also significant interaction with the subsurface
(Ni) layer.  The two surface states cross along the
$\overline{\Gamma J}$ direction bridging the minority gap between
the valence and the conduction band. Along the other directions
anticrossing occurs leading to band-gaps. Of interest are also the
saddle-like structures around the zone center which show up as van
Hove singularities in the DOS.

The  Ni terminated surface also shows two surface states, as can
be seen from figure~\ref{fig4}. In comparison to the  MnSb
termination, this  more open surface  leads  to rather  flat
dispersions of the surface states. Accordingly in the DOS, shown
in figure~\ref{fig3}, these surface states are much more intense
and effectively destroy the minority gap on this surface.

\subsection{Spin-polarization and magnetic moments}

\begin{figure}
\begin{center}
\includegraphics[angle=270,scale=0.23]{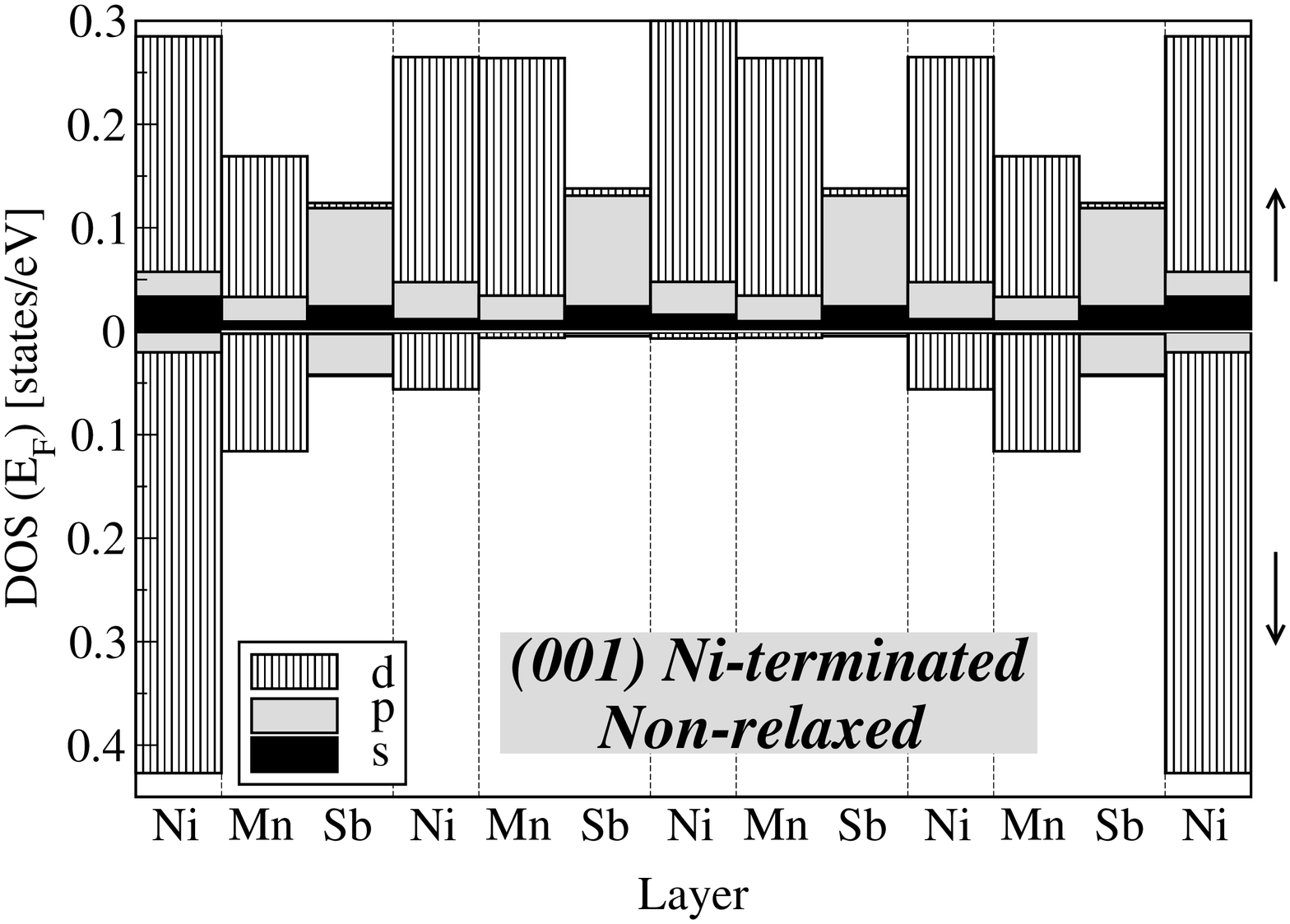}
\includegraphics[angle=270,scale=0.23]{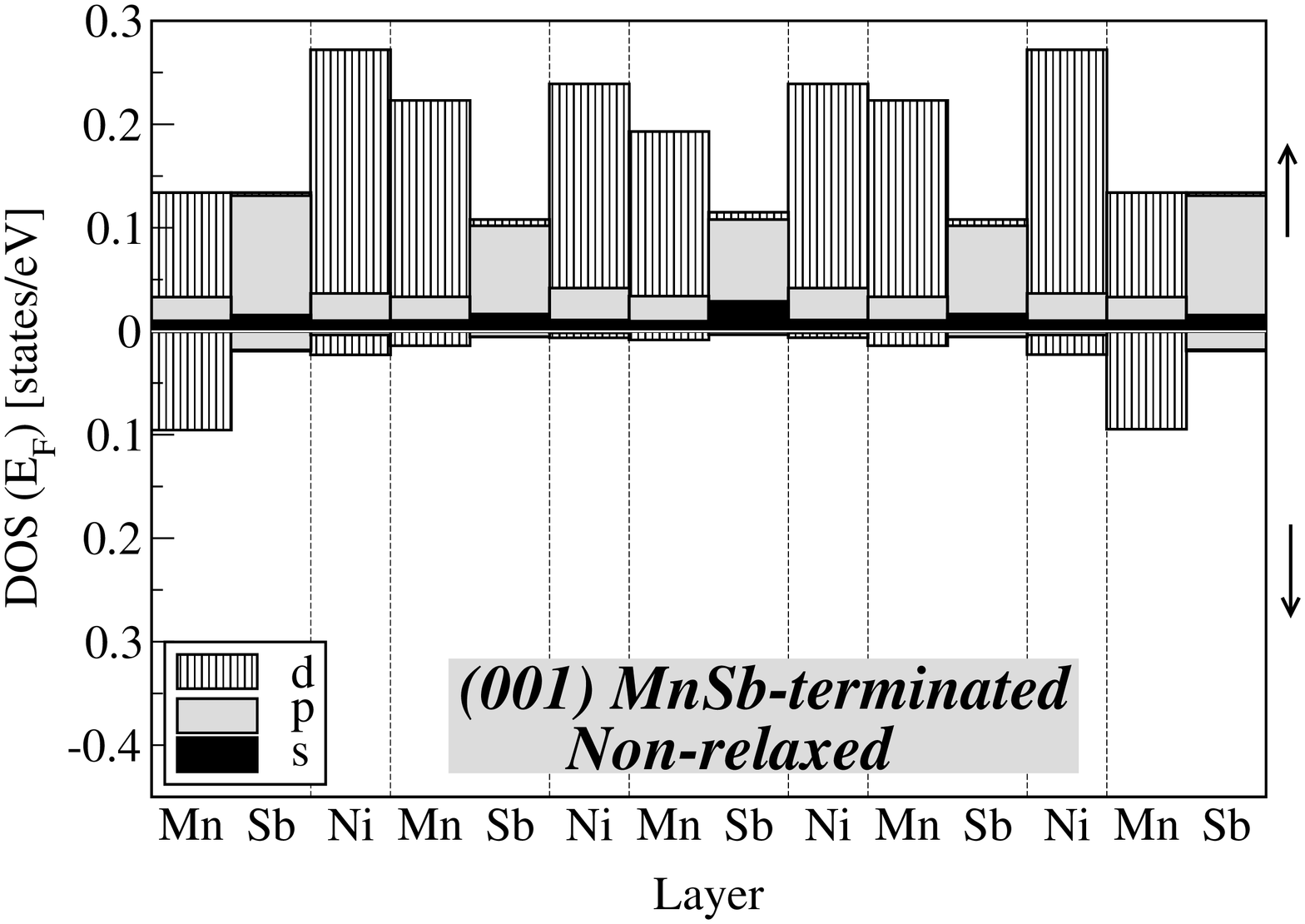}
\includegraphics[angle=270,scale=0.23]{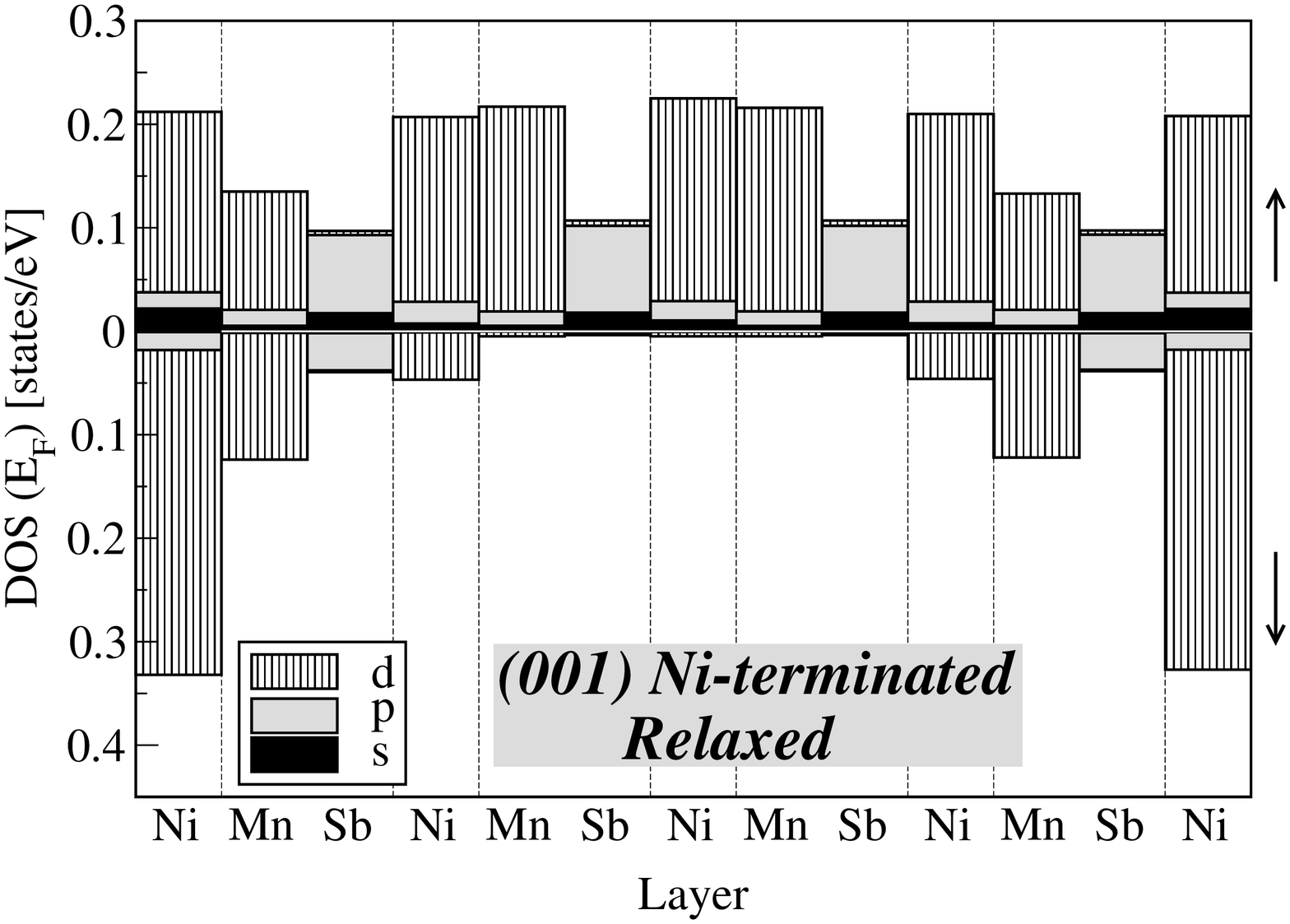}
\includegraphics[angle=270,scale=0.23]{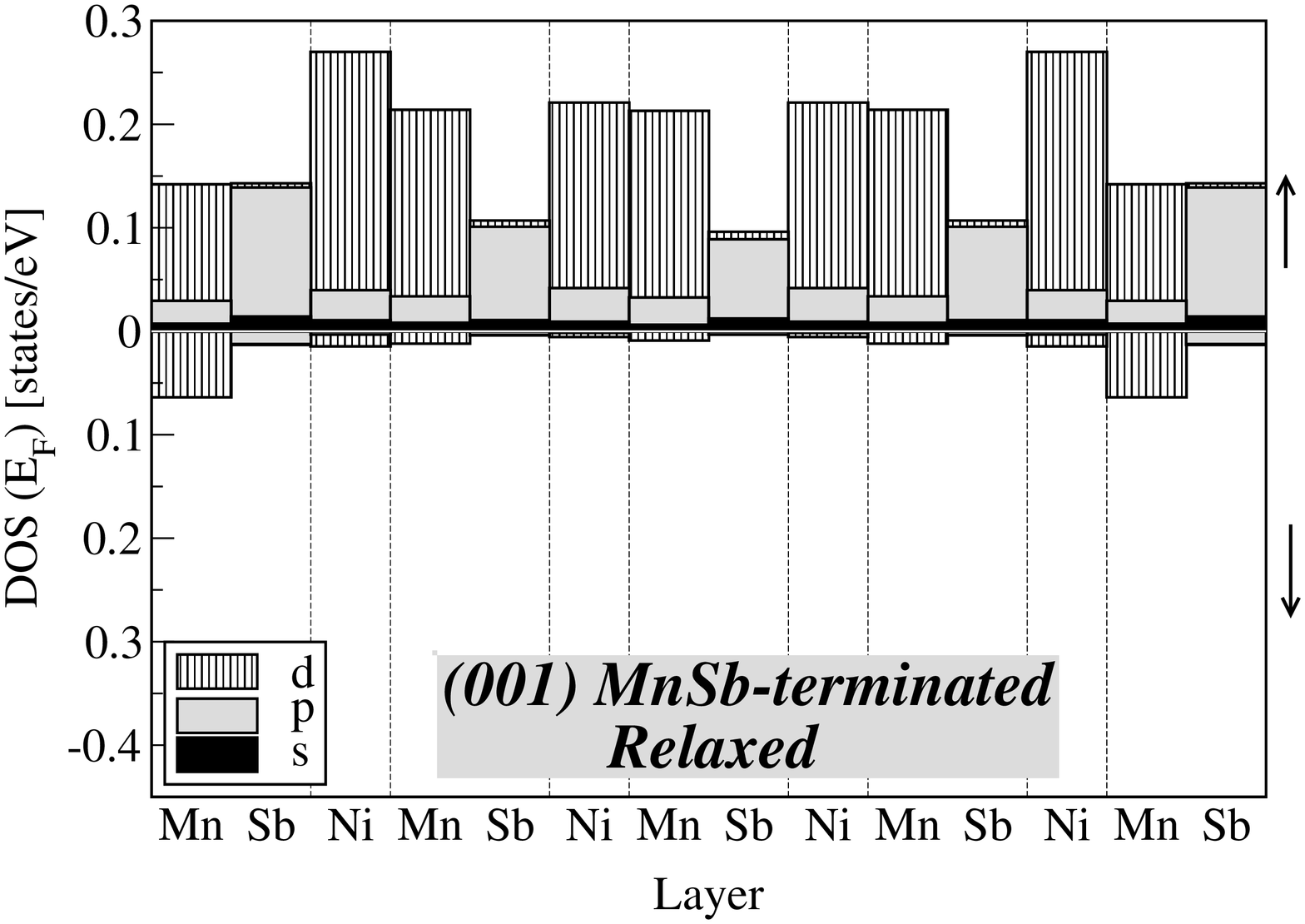}
\caption{Atom- and angular momentum-projected DOS at the Fermi level
for the different layers of the unrelaxed (top) and relaxed (bottom) film
with Ni (left) and MnSb (right) termination. Note, that the films in
the case of the (001) surfaces are inversion-symmetric.} \label{fig5}
\end{center}
\end{figure}
Using the above results, the spin-polarization at the Fermi level
can be determined. At interfaces, it is of prime importance since in a
current-injection experiment normally the electrons near the Fermi
level are involved \cite{Wunnicke}. In figure~\ref{fig5} we have
gathered the angular-momentum spin- and layer-projected DOS at the
Fermi level ($n^\uparrow (E_\mathrm{F})$ or $n^\downarrow
(E_\mathrm{F})$) for all the calculated films. The layers near the
edges of each figure represent the two equivalent surfaces
while the layers at the middle are bulk-like.  In the case of the
unrelaxed surfaces (upper panels) the minority DOS of the layers
at the middle of the film is almost zero and thus the thickness
of the film used in the calculation is sufficient to
realistically represent the real surface. Only in the case of the
relaxated MnSb-terminated surface, the Mn atom at the middle
layer presents a very small DOS. As expected, the states at the
Fermi level are mainly of $d$ character for Mn and Ni and of $p$-type
for Sb for both spin-directions.

\begin{table}
\caption{Spin-projected DOS at the Fermi level ($n^\uparrow
(E_\mathrm{F})$ or $n^\downarrow (E_\mathrm{F})$) for different
(001) surfaces taking into account either the top two layers, S
and S$-1$ (upper panel), or the top four layers (lower panel). The
spin-polarization is defined as $P_=\frac{n^\uparrow
(E_\mathrm{F})-n^\downarrow (E_\mathrm{F})}{n^\uparrow
(E_\mathrm{F})+n^\downarrow (E_\mathrm{F})}$ } \label{table1}
\begin{indented}
 \item[]
 \begin{tabular}{ll|rr|rr}\br
\multicolumn{2}{l|}{Spin} & (001) Ni  & (001) Ni  & (001) MnSb &
 (001) MnSb  \\ \multicolumn{2}{l|}{Polarisation} &  unrelaxed &  relaxed &  unrelaxed &
  relaxed \\ \mr
   Layers  & $n^\uparrow (E_\mathrm{F})$   & 0.855 &0.641 & 0.777  & 0.796   \\
  S, S$-1$   & $n^\downarrow (E_\mathrm{F})$ & 0.655 &0.556 & 0.161  & 0.107   \\
           & $P_1$                &  13\% & 7\%  & 66\%   & 76\%    \\  \mr
    Layers & $n^\uparrow (E_\mathrm{F})$   & 1.781 & 1.352& 1.573  & 1.543   \\
    S, S$-1$ & $n^\downarrow (E_\mathrm{F})$ & 0.730 & 0.618& 0.194  & 0.135   \\
  S$-2$, S$-3$ & $P_2$                &  42\% & 37\% & 78\%   &  84\%
\\ \br
\end{tabular}
\end{indented}
\end{table}

To expand our conclusions we need to quantify the DOS at the
Fermi level and in table \ref{table1} we have gathered the results
for all surfaces. We have calculated the spin-polarization either
taking into account only the first two surface layers $P_1$ or the
first four surface layers $P_2$. $P_2$ represents quite well the
experimental situation as the spin-polarization in the case of
films is usually measured by inverse photoemission which probes
the first few surface layers of the sample \cite{Borca02}. As expected,
the inclusion of more layers increases the spin-polarization since
deeper layers are more bulk-like. Relaxation in the case of the
Ni-terminated surface decreases the spin-polarization while in the
case of the MnSb-terminated surface the spin-polarization is
increased by the relaxation of the atomic positions.

In the case of the Ni terminated surface, the minority-spin DOS at
the Fermi level is quite large with respect to the majority DOS
and net polarization $P_2$ is 42\% for the unrelaxed case and
slightly decreases to 37\% by structural optimization. In the
case of the MnSb terminated surface the spin-polarization is much
larger  and now $P_2$ reaches a value of 84\% for the relaxed
structure, which means that more than 90\% of electrons at the
Fermi level are of majority-spin character. Our values for $P_1$
can be compared to reference \cite{GalanakisSurf001}, where
$\sim$ 0\% and 38 \% spin-polarisation have been found for the Ni
and MnSb terminations, respectively.
As can be seen from figure~\ref{fig5} the
main difference between the two different terminations is the
contribution of the Ni spin-down states. In the case of the MnSb
surface the Ni in the subsurface layer has a negligible  DOS at
the Fermi level with respect to  the Ni-terminated surface. It is
interesting also to see from this figure, that for both
terminations the spin-polarization of the Mn near the surface at
the Fermi level is close to zero while Sb atoms in both cases
show a large spin-polarization.  The calculated $P_2$ value of
84\% for the MnSb terminated surface is larger  than the
experimental value of 67\% obtained by Ristoiu and collaborators
\cite{Ristoiu} for a thin-film terminated in a MnSb stoichiometric
alloy surface layer. A direct comparison between experiment and
theory is not straightforward, since experimentally different
layers will contribute with different weight to the spin-polarization.

\begin{table}
\caption{Atom-projected spin magnetic moments ($m_{spin}$) in
$\mu_B$ for the atoms at the top four layers for both Ni- and
MnSb-terminated (001) surfaces for both relaxed and
unrelaxed cases.} \label{table2}
\begin{indented}
 \item[]
\begin{tabular}{llrr|llrr} \br
  \multicolumn{4}{c|}{ Ni-terminated} &  \multicolumn{4}{c}{ MnSb-terminated} \\ 
         &    & unrel. & rel.  &      &        & unrel. & rel.     \\ \mr 
 Ni & (S)   & $ 0.44 $   &$ 0.38 $ &Mn & (S)      & $ 3.94 $   &$ 3.93 $ \\ 
 Mn & (S$-1$) & $ 3.79 $   &$ 3.64 $ &Sb & (S)      & $-0.10 $   &$-0.10 $ \\ 
 Sb & (S$-1$) & $-0.04 $   &$-0.04 $ &Ni & (S$-1$)    & $ 0.21 $   &$ 0.24 $ \\ 
 Ni & (S$-2$) & $ 0.27 $   &$ 0.28 $ &Mn & (S$-2$)    & $ 3.66 $   &$ 3.69 $ \\ 
 Mn & (S$-3$) & $ 3.71 $   &$ 3.59 $ &Sb & (S$-2$)    & $-0.07 $   &$-0.07 $ \\ 
 Sb & (S$-3$) & $-0.06 $   &$-0.05 $ &Ni & (S$-3$)    & $ 0.27 $   &$ 0.24 $  
\\ \br
\end{tabular}
\end{indented}
\end{table}
In table~\ref{table2} we  have gathered  the spin magnetic moments
of the atoms in the surface and subsurface layers. We should
notice that relaxation has in most cases only a small effect on the spin
moments. Even for the surface layer which shows the largest
relaxation effects, spin moments change by at most $0.06 \mu_B$.

In the case of the MnSb terminated NiMnSb(001) surface, the
surface layer loses $\sim$0.3$e^-$ as compared to the bulk. This
is due to the spilling out of charge into the vacuum and affects
mainly the Mn minority-spin electrons. Therefore, the Mn's spin
magnetic moment increases with respect to the bulk and is slightly
more than $3.9 \mu_B$. This behaviour arises from the reduced
symmetry of the Mn atom in the surface which loses two of the four
neighbouring Ni atoms. In the majority band this leads to a
narrowing of the $d$-DOS and this affects also the subsurface Ni
layer that loses 0.1 $e^-$, while in the minority valence band the
Mn $d$-contribution decreases by 0.2 $e^-$. Moreover, the
splitting between the unoccupied Mn states above $E_\mathrm{F}$
and the center of the occupied Mn states decreases and at
$E_\mathrm{F}$ a surface states appears. We  should also mention
here that in the case of a half-metallic material the total spin
magnetic moment per unit cell should be an integer since the
number of both the total valence electrons and the minority-spin
occupied states are integers; the spin moment in $\mu_B$ is simply
the number of uncompensated spins, i.e.\ $4 \mu_B$. In the case of
the surfaces the half-metallic character is lost and an increase
of the total spin moment is observed, which is no more an integer
number.

In the case of the Ni terminated surface, the changes in the DOS
compared to the bulk are more pronounced. The Ni atom in the
surface loses some charge.  As was the case for the Mn surface atom
in the MnSb terminated surface, also the surface atoms  spin
magnetic moment is increased
(see table \ref{table2}). The Mn and Sb atoms in the subsurface
layer present a charge transfer comparable to the bulk compound
and also a comparable spin moment.

\section{(111) surfaces} \label{sec5}

\subsection{Structure and relaxation}

\begin{table}
\caption{Relative changes in the distance $\Delta d_{ij}$ between
successive layers $i,j$ when the atomic positions were relaxed for
the (111) surfaces. Negative signs correspond to contractions,
positive to expansions.
} \label{table3}
\begin{indented}
 \item[]
 \begin{tabular}{l|rrr} \br
              & $\Delta d_{12}$ & $\Delta d_{23}$ & $\Delta d_{34}$ \\ \mr
 Ni-Sb-Mn-... & $ -23$\% & $ 2$\% & $< 1$\%  \\
 Ni-Mn-Sb-... & $ -18$\% & $ 4$\% & $ -3$\%  \\
 Mn-Ni-Sb-... & $ -13$\% & $-5$\% & $  2$\%  \\
 Mn-Sb-Ni-... & $ -16$\% & $18$\% & $ \sim 0$\%  \\
 Sb-Ni-Mn-... & $   2$\% &$-11$\% & $  4$\%  \\
 Sb-Mn-Ni-... & $ -16$\% & $32$\% & $ -7$\%  \\ \br
\end{tabular}
\end{indented}
\end{table}
In this section,  we will study the (111)
surfaces of NiMnSb. Relaxations in the case of the (111) surfaces
are considerably larger than for the (001) ones. In table
\ref{table3} we have gathered the change in the distance between
two successive layers with respect to the unrelaxed cases. The
closed packed (111) layers contain only one chemical element.
Note, that in the unrelaxed cases the distance between Sb and Mn
successive layers is twice the distance between a Ni and a Mn or
Sb layer.

When the (111) surface is Ni-terminated, the Ni atoms at the
surface layer move closer to the subsurface layer and the
contraction is 23\% and 18\% for Sb and Mn as subsurface layers,
respectively. Relaxations are much less important deeper than the
surface layer. When the surface is Mn terminated with a Mn-Ni-Sb-...
stacking sequence, the Mn atoms move closer to Ni  due to the lower
coordination. In the case of a  Mn-Sb-Ni-... sequence, relaxations
are more important since the Mn-Sb distance is twice the Mn-Ni one.
From table \ref{table3} we see, that this results not only in a
large contraction of the first two layers (negative $\Delta d_{12}$),
but also in a expansion of the next interlayer distance, $\Delta d_{23}$.
A similar effect can be observed on the Sb-Mn-Ni-... terminated
surface.
In the latter case, similar relaxations have also been obtained
by Jenkins~\cite{Jenkins04}

\subsection{Density of states and bandstructures}

\begin{figure}
\begin{center}
\includegraphics[angle=270,width=\textwidth]{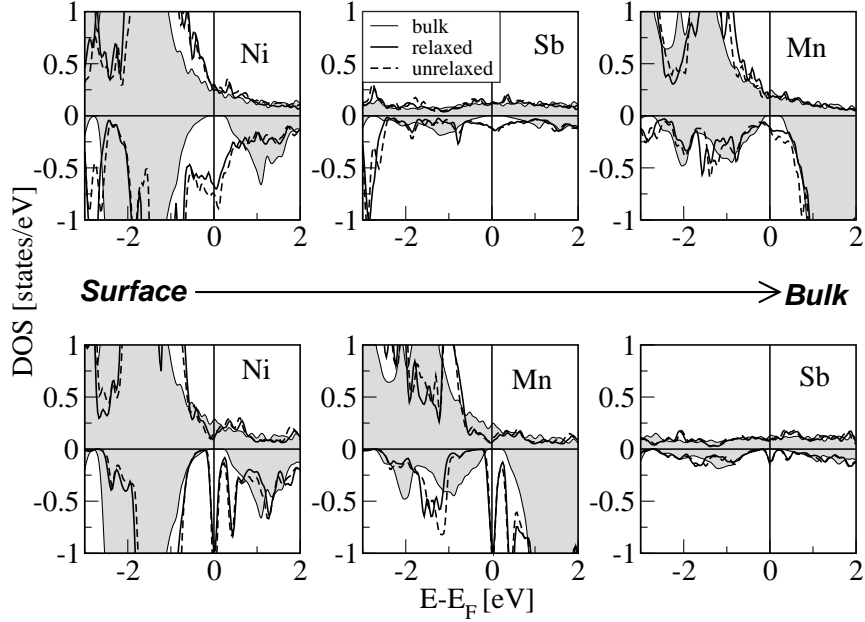}
\caption{Same as figure~\ref{fig3} for the Ni-terminated (111)
surfaces. There are two different Ni terminations, either with a
Sb or a Mn layer as the subsurface one.} \label{fig6}
\end{center}
\end{figure}
\begin{figure}
\begin{center}
\includegraphics[scale=0.67]{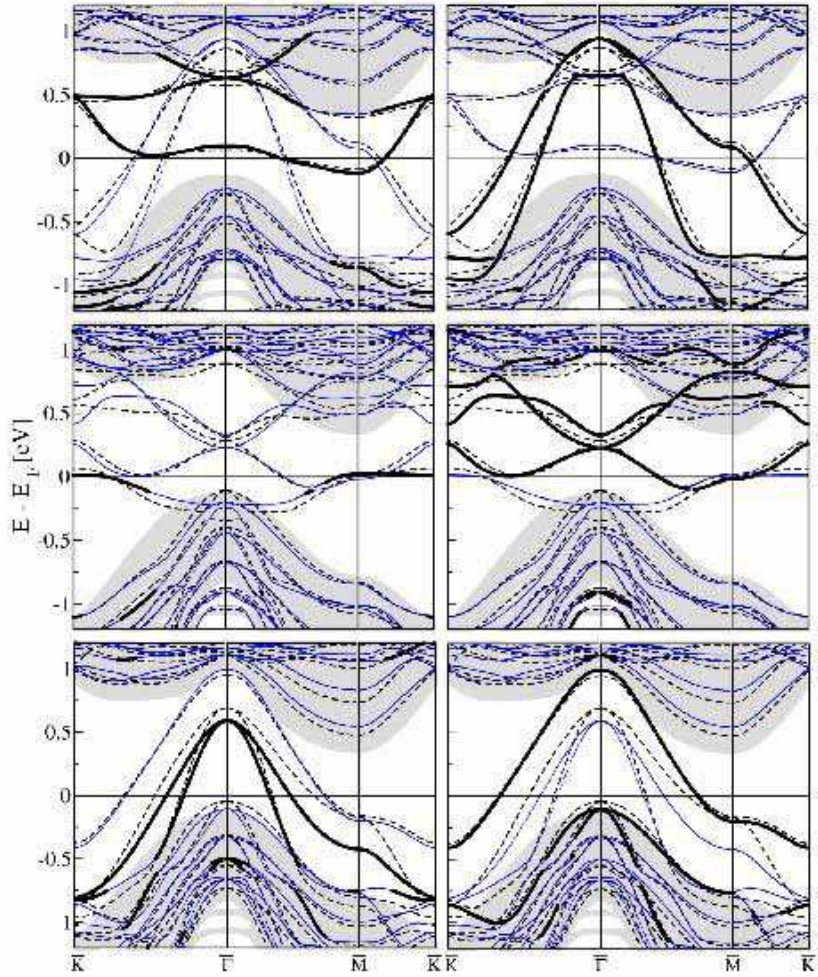}
\caption{Surface bandstructures for the Ni (top) Mn (middle) and
Sb (bottom) terminated (111) oriented NiMnSb films. The two surfaces
of the film have different subsurface layers and, therefore, give
rise to different surface states. The left column shows the surface
states of the Ni-Mn-Sb-..., Mn-Ni-Sb-... and Sb-Mn-Ni-... terminated
surfaces (top to bottom), while on the right the states from the
Ni-Sb-Mn-..., Mn-Sb-Ni-...  and Sb-Ni-Mn-... surfaces are marked. Otherwise
the labeling is identical to figure~\ref{fig4}.} \label{fig7}
\end{center}
\end{figure}

In figure~\ref{fig6} we have gathered the spin-resolved density of
states (DOS) for the three layers closest to the surface for both
types of Ni termination. For the  Ni-Mn-Sb-... termination, there
is a minority surface state pinned exactly at the Fermi level
which completely destroys the half-metallicity.
The population of the majority states increases and due to
the exchange splitting the minority states are pushed higher in
energy. This results in a very sharp shape of the surface state.
Actually there are two surface states as we will discuss later in
this paragraph. This phenomenon is more pronounced for the Mn atom
at the subsurface layer, whose occupied minority states have a
small weight,  and thus it presents a much larger exchange
splitting energy since this one scales with the spin magnetic
moment. This surface state gradually decays  and for the Ni atom
at the S$-3$ position (not shown here) it practically vanishes. We
can identify this surface states also in the surface
band-structure presented in the upper left of figure~\ref{fig7}. Thick lines mark
the surface states for this termination. We observe now two
surface states, similar to the (001) surfaces, which are very
narrow-spread in energy around the Fermi level, resulting in a
very sharp peak structure.

In the case of the Ni-Sb-Mn-... surface, the Ni bands even
move slightly higher in energy. Therefore, the Ni spin moment is much
smaller and the Mn atom is deep in the substrate. The surface states
are now much more extended  in the energy axis and cannot be well
separated from the rest of the DOS as shown in figure~\ref{fig6}.
This situation is similar to the Ni terminated (001) surface.
These surfaces states (figure~\ref{fig7})
are clearly much broader in energy than the states
in the case of the Ni-Mn-Sb-... termination resulting in a very
extended peak at the Fermi level which is not easily distinguished
in the DOS. Our band-structure is similar to the one calculated by
Jenkins \cite{Jenkins04}.

\begin{figure}
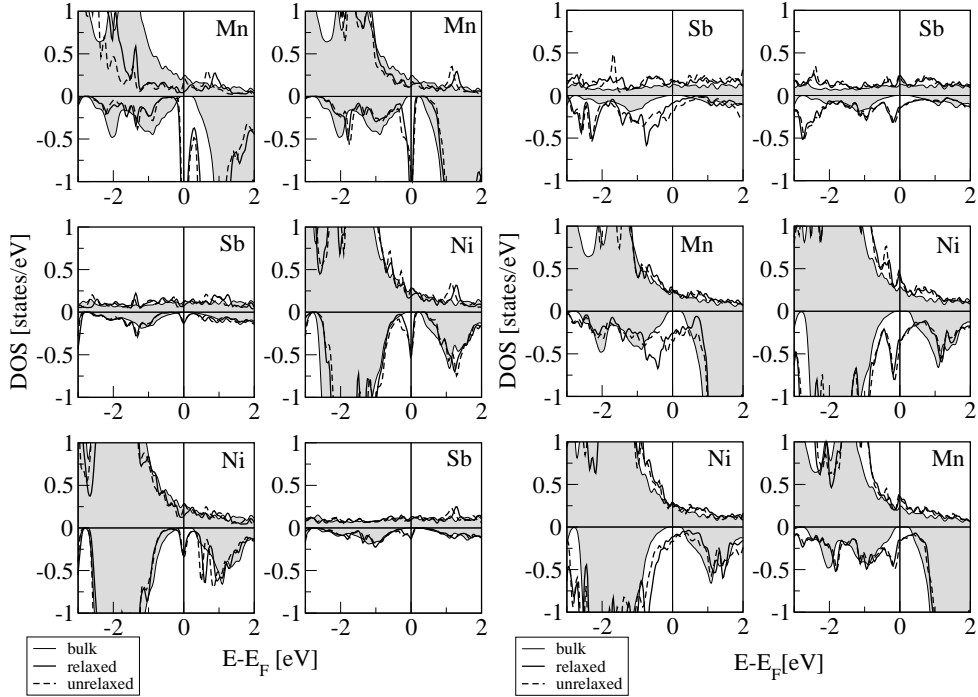

\begin{center}
\includegraphics[scale=0.35]{fig8a.eps}
\includegraphics[scale=0.35]{fig8b.eps}
\caption{Same as figure~\ref{fig6} for the Mn- and Sb-terminated
(111) surfaces. Each column represents a different surface
termination. The top panels represent the surface layers, the
middle ones the subsurface layers and the bottom panels the
subsubsurface ones.} \label{fig8}
\end{center}
\end{figure}
In figure~\ref{fig8} we have gathered the DOS for the
first three layers for all Mn and Sb surfaces. In the case of the
Mn-terminated surfaces, there is a minority surface state pinned
exactly at the Fermi level which destroys the half-metallicity and
which is also visible in the Ni subsurface layer, but vanishes
in the next Mn layer (not shown here). The overall DOS are similar
to the bulk case and the increase of the Mn spin moment at the
interface reflects that high-lying majority antibonding
$d$-states, which are above the Fermi level in the bulk but now
move below it, push also the majority bands somewhat lower in
energy \cite{GalanakisBulk}. The surface states in the reciprocal
space are similar to the Ni-Mn-Sb-... case shown in figure~\ref{fig7}.
In the case of the Mn-Sb-Ni-... surface, there are three surface
states with very flat dispersion while in the case of the
Mn-Ni-Sb-... (111) surface there is just one very flat surface
state centered along the $\overline{MK}$ line leading to the very sharp
peak shown in figure~\ref{fig8} and leaving a band gap just above
the Fermi level. Also in the case of the Sb terminated surface
there is a minority surface state slightly below the Fermi level
which also destroys the half-metallicity at the surface. Its
intensity is large also for the Ni at the subsurface layer but
already for the Mn atom it starts to smear out. These surface
states can also be traced in the reciprocal space
as shown in the lower panels of figure~\ref{fig7}. They are very
wide in energy and thus can not be well separated from the rest of
the DOS. Where comparable, our surface band-structures agree with
the results of Jenkins \cite{Jenkins04}.

\subsection{Spin-polarization and magnetic moments}

\begin{table}
\caption{Atom-projected spin magnetic moments ($m_{spin}$) in
$\mu_B$ for the atoms at the top six layers for all Ni-, Mn- and
Sb-terminated (111) surfaces and for both relaxed and unrelaxed
cases.} \label{table4}
\begin{indented}
 \item[]
 \begin{tabular}{lrr|lrr|lrr} \br
 \multicolumn{3}{c|}{ Ni-Mn-Sb-...} & \multicolumn{3}{c|}{ Mn-Ni-Sb-...} & \multicolumn{3}{c}{ Sb-Mn-Ni-...} \\
            &  unrel. & rel.  &           & unrel. & rel. &             & unrel. & rel.  \\ \mr
Ni(S)   & $  0.54 $  & $ 0.47 $ & Mn(S)   & $  3.90 $  & $ 3.73 $ &Sb(S)     & $-0.19 $  & $-0.21 $ \\
Mn      & $  3.89 $  & $ 3.77 $ & Ni      & $  0.23 $  & $ 0.28 $ &Mn        & $ 3.62 $  & $ 3.56 $ \\
Sb      & $ -0.05 $  & $-0.05 $ & Sb      & $ -0.07 $  & $-0.07 $ &Ni        & $ 0.19 $  & $ 0.09 $ \\
Ni      & $  0.27 $  & $ 0.27 $ & Mn      & $  3.63 $  & $ 3.56 $ &Sb        & $-0.07 $  & $-0.07 $ \\
Mn      & $  3.70 $  & $ 3.57 $ & Ni      & $  0.24 $  & $ 0.25 $ &Mn        & $ 3.65 $  & $ 3.59 $ \\
Sb      & $ -0.05 $  & $-0.05 $ & Sb      & $ -0.07 $  & $-0.08 $ &Ni        & $ 0.26 $  & $ 0.24 $ \\
\br
 \multicolumn{3}{c|}{ Ni-Sb-Mn-...} & \multicolumn{3}{c|}{ Mn-Sb-Ni-...} & \multicolumn{3}{c}{ Sb-Ni-Mn-...} \\
            &  unrel. & rel.  &           & unrel. & rel. &             & unrel. & rel.  \\ \mr
Ni(S)   & $ 0.30  $  & $ 0.30 $ & Mn(S)   & $  4.16 $  & $ 3.89 $  & Sb(S)   & $-0.12 $  & $-0.12 $ \\
Sb      & $-0.04  $  & $-0.04 $ & Sb      & $ -0.04 $  & $-0.05 $  & Ni      & $ 0.13 $  & $ 0.15 $ \\
Mn      & $ 3.71  $  & $ 3.54 $ & Ni      & $  0.30 $  & $ 0.33 $  & Mn      & $ 3.52 $  & $ 3.50 $ \\
Ni      & $ 0.23  $  & $ 0.21 $ & Mn      & $  3.70 $  & $ 3.66 $  & Sb      & $-0.06 $  & $-0.07 $ \\
Sb      & $-0.07  $  & $-0.07 $ & Sb      & $ -0.06 $  & $-0.07 $  & Ni      & $ 0.26 $  & $ 0.24 $ \\
Mn      & $ 3.68  $  & $ 3.56 $ & Ni      & $  0.27 $  & $ 0.27 $
& Mn      & $ 3.70 $  & $ 3.69 $ \\ \br
\end{tabular}
\end{indented}
\end{table}

In the bulk case Ni has four Mn and four Sb atoms as first
neighbours. At the Ni terminated (111) surface, the Ni atom at the
surface  loses four out of its eight first neighbours. In the
case of the Ni-Mn-Sb-... termination it loses three Sb atoms and one
Mn atom while in the Ni-Sb-Mn-... case one Sb and three Mn atoms.

In table \ref{table4} we have gathered the spin moments for the
first six layers for all surfaces under study. Generally, we can
observe that relaxations tend to decrease the Mn moments, while in
some cases the Ni or Sb moments can increase slightly. In the case
of the Ni-Mn-Sb-... termination, both Ni and Mn atoms at the
surface have very large moments with respect to both the bulk
calculations and the Ni-Sb-Mn-... case. Especially the Ni moment
is almost doubled ($0.47 \mu_B$) with respect to the bulk value of
$0.25 \mu_B$. In the bulk NiMnSb the minority gap is
created by the hybridization between the $d$-orbitals of the Ni
and Mn atoms, but the Sb atom plays also a crucial role since it
provides states lower in energy than the $d$ bands which
accommodate electrons of   the transition metal atoms
\cite{GalanakisBulk}.  At the Ni-Mn-Sb-... terminated surface,
each Ni surface atoms loses three out of the
four Sb first neighbours and they regain most of the charge
accommodated in the  $p$-bands of Sb. These extra electrons fill
up mostly majority states,  increasing the Ni spin moment. Mn spin
moment is also increased since Mn and Ni majority $d$-states
strongly hybridize forming a common majority band as it was shown
in reference \cite{GalanakisBulk}. Thus the spin moment of Mn at
the subsurface layer increases to $3.77 \mu_B$ ($3.89 \mu_B$ in
the unrelaxed case) with respect to the bulk value of $3.73
\mu_B$. If one goes further away from the surface, the atoms have
a bulklike environment and their spin moments are similar to the
bulk moments. In the Ni-Sb-Mn-... surface, Ni at the surface loses
only one Sb first neighbour and the effect of the cut-off
neighbours is much smaller. The moment is slightly smaller than
the bulk one mainly due to a surface state at the minority band
shown in figure~\ref{fig6}. Already the Sb subsurface atom regains
a bulklike behaviour for the spin moment.

In the case of the Mn surfaces, Mn at the surface layer loses half
of its Sb second neighbours. Similarly to what happened in the
case of the Ni-Mn-Sb-... surface, its spin moment is strongly
enhanced especially in the  Mn-Sb-Ni-... case (to $3.89 \mu_B$).
In this case, we can think that  Mn has a subsurface layer made up
by voids and thus the hybridization between the Mn $d$-orbitals
and the Sb $p$- and Ni $d$-orbitals is strongly reduced leading to
an increase of its spin moment with respect to the Mn-Ni-Sb-...
case. Relaxations tend to decrease the Mn moment, but due to the
large increase of $\Delta d_{23}$ in the  Mn-Sb-Ni-... case, the
Ni moment increases here.  The atoms deeper in the surface quickly
reach a bulklike behaviour.

Following the same arguments as for Mn, one can understand also
the behaviour  of the spin moments for the Sb terminated surfaces
presented  in table \ref{table4}. The absolute value of the Sb
spin moment at the surface layer increases with respect to the
bulk. When the subsurface layer is a ``void layer'' (Sb-Mn-Ni-..
case), the hybridization effects are less important and the Sb
spin moment can reach a value of $-0.2 \mu_B$. This is almost three times
the bulk value of $-0.07 \mu_B$ and double the value for the (001)
surface of $-0.1$ $\mu_B$. The change in the Sb $p$-bands
influences also through hybridization the bands of the transition
metal atoms for which now the minority bands population increases
leading to smaller spin moments of the Ni and Mn atoms at the
subsurface layers. The phenomenon is more intense in the case of
Sb-Ni-Mn-... where the Ni layer is just below the Sb surface layer
and the reduction in the spin moments of Ni and Sb is much larger
than in the Sb-Mn-Ni-... case.

\begin{table}
\caption{
Spin-polarization at the Fermi level for different (111) surfaces taking into
account either the top three layers ($P_1$) or the top six layers
($P_2$). The spin-polarization is defined as in table \ref{table1}.
} \label{table5}
\begin{indented}
 \item[]
 \begin{tabular}{l|rr|rr}\br
             & \multicolumn{2}{c|}{$P_1$} & \multicolumn{2}{c}{$P_2$} \\
             &  unrelaxed & relaxed   &   unrelaxed & relaxed  \\ \mr
Ni-Sb-Mn-... &$ -13$\% &$ -19$\% &$  17$\% &$   8$\% \\
Ni-Mn-Sb-... &$ -70$\% &$ -74$\% &$ -49$\% &$ -67$\% \\
Mn-Ni-Sb-... &$ -64$\% &$ -66$\% &$ -16$\% &$ -49$\% \\
Mn-Sb-Ni-... &$ -52$\% &$ -78$\% &$ -31$\% &$ -58$\% \\
Sb-Mn-Ni-... &$   9$\% &$  29$\% &$  35$\% &$  49$\% \\
Sb-Ni-Mn-... &$  20$\% &$  28$\% &$  33$\% &$  42$\% \\
 \br
\end{tabular}
\end{indented}
\end{table}
Finally, we have collected the spin-polarization at the Fermi
level in table \ref{table5}. Since we have seen that only the Sb
terminated surfaces have surface states that are not localised too
narrow around the Fermi level, we observe only on these surfaces
substantial spin-polarizations. In these cases, relaxations have
the effect to increase the spin-polarization at the Fermi level,
while in most other cases large negative values are observed,
which get even more negative when structural relaxations are taken
into account.

\section{Summary} \label{sec6}

We have performed \textit{ab-initio} calculations based on the
full-potential linearised augmented plane-wave method for the (001)
and (111) surfaces of the half-metallic NiMnSb Heusler alloy. The
MnSb terminated (001) surfaces  present electronic and magnetic
properties similar to the bulk compounds. There is however a small
finite Mn-$d$ and Sb-$p$ DOS within the bulk spin-down gap and
these surface states are strongly localised at the surface layer.
The spin-polarization at the Fermi level for this termination
reaches the 84\%. The (001) surfaces terminated at Ni present a
quite large density of states at the Fermi level and properties
considerably different from the bulk and the MnSb terminated
surfaces. In both terminations, two distinct surface states
can be seen in the surface bandstructure, which are of quite
different character on the two surfaces.

In all (111) surfaces minority-spin surface states destroy the
half-metallicity at the surface. They are pinned at the Fermi
level for the Ni and Mn terminated surfaces but are slightly below
the Fermi level for the Sb terminated ones. They are localised close
to the surface region and typically vanish within few atomic layers.
Surface states show a variety of dispersion relations as was shown
by the surface band-structures. In the case of the Ni
surface with Mn as subsurface layer, Ni-Mn-Sb-..., the loss of
three out of the four Sb first neighbours leads to a doubling of
the Ni spin moment while in the Ni-Sb-Mn-... case it is near the
bulk value. For the Mn and Sb terminations the lowering of the
coordination increases the surface spin moments and the
enhancement is larger when the subsurface layer is not a Ni one.
Only on the Sb terminated surfaces substantial spin-polarizations
can be observed, specifically when relaxations are taken into account.


\ack{This work was financed in part by the BMBF under auspices of
the Deutsches Elektronen-Synchrotron DESY under contract no. 05
KS1MPC/4.}

\end{document}